\documentstyle[12pt%
,epsfig%
,cite%
]{article}
\def\Rf#1#2#3#4{{#1} {#2} (19#3) #4}
\def\Rfi#1#2#3#4#5{{#1} #2 {#3} (19#4) #5}

\def\IJMP{{Int. J. Mod. Phys.} A}

\def\PL{{Phys. Lett.}  B}
\def\PRp{Phys. Reports}
\def\PR{Phys. Rev.}
\def\ZP{{Z. Phys.} C}
\def\YaF{Sov. J. Nucl. Phys.}

\def\ib{\em ibid.}
\def\ea{\em et al.}
\def\ct{\cite}
\def\fn{\footnote}
\def\np{\newpage}

\def\be{\begin{equation}}
\def\ee{\end{equation}}
\def\bea{\begin{eqnarray}}
\def\eea{\end{eqnarray}}
\def\la{\label}
\def\vs{\vspace}
\def\bi{\bibitem}

\def\rT{\rm T}
\def\rd{\rm d}
\def\e{\eta}
\def\ec{\e_{\rm 0}}
\def\ve{\varepsilon}
\def\vt{\vartheta}
\def\ef{\stackrel{\sim}{\e}}
\def\ecf{\ef_{\rm 0}}
\def\al{\langle}
\def\ar{\rangle}
\def\De{\Delta\e}
\def\de{\delta\e}
\def\dep{\delta\ef}
\def\dn{\delta n}
\def\r{\rho}
\def\ddf{\r(\ef)}
\def\mn{\rm min}
\def\mx{\rm max}
\def\se{\simeq}
\def\c2{\chi^2/{\rm DOF}}
\def\st{(\rm stat)}
\def\sy{(\rm syst)}
\def\rmx{\r_{\mx}}
\def\rmxa{\al \rmx \ar }
\def\ma{14<n<20}
\def\mb{24<n<30}
\def\rms{{\rm RMS}}
\def\aln{all $n$}

\def\wea{E.K. Sarkisyan {\ea},}

\begin{document}

\pagestyle{empty}
%%%%%%%%%%%%%%%%%%%%%%%%%%%%%%%%%%%%%%%%%%%%%%%%%%
%                                                %
%                  TITLE LIST                    %
%                                                %
%%%%%%%%%%%%%%%%%%%%%%%%%%%%%%%%%%%%%%%%%%%%%%%%%%
%
\begin{center}
{\Large \bf Study of multiparticle spikes\\
in central 4.5$A$ GeV/$c$ C-Cu collisions}
\bigskip

{\large G.L. Gogiberidze$^{\rm a,}$\fn {On leave from Institute of
Physics, Tbilisi 380077, Georgia.}, L.K. Gelovani$^{\rm a,1}$,
E.K. Sarkisyan$^{\rm b}$} \\
\medskip

$^{\rm a}$~{\it Joint Institute for Nuclear Research, P.O.B. 79,
Moscow 101000, Russia}

$^{\rm b}~${\it School of Physics \& Astronomy,
Tel Aviv University, Tel Aviv 69978, Israel}

\vs*{1.8cm}
%%%%%%%%%%%%%%%%%%%%%%%%%%%%%%%%%%%%%%%%%%%%%%%%%%%%%%%%%%%%%%
\end{center}
\medskip
\centerline{\large \bf Abstract}
\vs*{1.cm}

\noindent
An analysis of local fluctuations, or spikes, is performed for charged
particles produced in central C-Cu collisions at 4.5 GeV/$c$/nucleon. The
distributions of spike-centers and the maximum density distributions are
investigated for different narrow pseudorapidity windows to search for
multiparticle dynamical correlations. 
Two peaks over statistical background are observed in the spike-center
distributions with the structure similar to that expected from the
coherent gluon radiation model and recently found in hadronic
interactions. The dynamical contribution to maximum density fluctuations
are obtained to be hidden by statistical correlations, though behavior of
the distributions shows qualitative agreement with that from the
one-dimensional intermittency model. 
The observed features of the two different approaches, coherent vs. 
stochastic, to the formation of the local dynamical fluctuations are
discussed.

%%%%%%%%%%%%%%%%%%%%%%%%%%%%%%%%%%%%%%%%%%%%%%%%%%
%                                                %
%     END OF THE TITLE LIST                      %
%                                                %
%%%%%%%%%%%%%%%%%%%%%%%%%%%%%%%%%%%%%%%%%%%%%%%%%%

\np
\pagestyle{plain}
\setcounter{page}{1}

%%%%%%%%%%%%%%%%%%%%%%%%%%%%%%%%%%%%%%%%%%%%%%%%%%
%                                                %
%    BEGINNING OF TEXT                           %
%                                                %
%%%%%%%%%%%%%%%%%%%%%%%%%%%%%%%%%%%%%%%%%%%%%%%%%%

\section{Introduction}

Reflecting an intermittent structure of distributions of particles
produced in high-energy collisions, local fluctuations play an important
r\^ole in hadronization process.  Intermittency/fractality studies
\ct{rev1} show these fluctuations to be significant in investigations of
multihadronic production as well as to be a possible signal of quark-gluon
plasma formation.  Although a vast activity in study self-similarity have
disentangled contributions of different known mechanisms to production of
dynamical fluctuations/correlations, an origin of the latters remains
still unclear. This arises need for applying more direct (and additional) 
methods of studying properties of dense groups of particles.

The data used here has been considered in our recent studies of
self-similar nature of multihadronic production \ct{mn,mg}. Strong
multifractality has been found, pointing out a possible non-thermal phase
transition and two different regimes in particle production during the
cascade. This conclusion confirms results of our earlier studies
\ct{pl3,pl4} made with less statistics and has been recently supported by
similar observations reported for ultra-relativistic nuclear collisions
\ct{nt}. Further analysis \ct{mg} have shown more structure in dynamical
fluctuations indicating chaotic nature of multiparticle production with a
specific scaling-law referred to as erraticity \ct{er}. 

Although the self-similar underlying dynamics has been obtained in searches in
fractality terms, no dynamical effects was revealed in the investigations
of maximum density distributions \ct{y1}.  However, the analogous study
performed for hadron-emulsion data in the energy range of 200-400 GeV
\ct{mx} has indicated the existence of dynamical multiparticle correlations
and clustering seen for high-density spikes.

In this letter we investigate local pseudorapidity fluctuations, or
spikes, analyzing two types of distributions, assigned to coherent vs.
chaotic approaches in particle-emission process.  Respectively, the
distributions of centers of spikes and maximum density distributions are
considered, based on the methods, used earlier and shown to be
distinguished by the reliability and the high certainty in searching for
multiparticle dynamical correlations \ct{revd,mxt,rev3,rev4}. Study of
fluctuations in limited regions of collision phase-space allows reducing
of contribution from conservation constraints and elicits underlying
dynamics of particle-production process.

\section{Data sample and analysis procedure}

\subsection{Data sample}

The results presented are based on experimental data came from interactions of
the JINR Synchrophasotron (Dubna) 4.5~$A$ GeV$/c$ $^{12}$C beam with a copper
target inside the 2m Streamer Chamber SKM-200 \cite{skm}. A central collision
trigger was used: absence of charged particles with momenta $p>3$ GeV$/c$ in a
forward cone of 2.4$^{\circ}$ was required. A more detailed description of the
set-up design and data reduction procedure are given elsewhere \ct{skm,skm1}.
Systematic errors related to the trigger effects, low-energy pion and proton
detection, the admixture of electrons etc. does not exceed 3\% \ct{skm2,skm1}.

The scanning and the handling of the film data were carried out on special
scanning tables of the Lebedev Physical Institute (Moscow) \cite{obr}. The
average measurement error in the momentum $\al\ve_p/p\ar$ was about
12$\%$, and that in the polar angle measurements was
$\al\ve_{\vt}\ar\simeq 2^{\circ}$. In total, 663 events with charged
particles in the pseudorapidity range of $\De= 0.2-2.8$ (in the laboratory
frame) were processed.  The angular measurement accuracy does not exceed
0.1 in the $\e$-units. In addition, particles with $p_{\rT}>1$ GeV/$c$ are
excluded from the investigation as far as no negative charged particles
were observed with such a transverse momentum.  Under the assumption of an
equal number of positive and negative pions, this cut was applied to
eliminate the contribution of protons. After the kinematic cuts, the mean
multiplicity is $23.8\pm 0.4$. 

\subsection{Analysis procedure}

The multiparticle fluctuations are studied in the pseudorapidity
phase-space regions. The following procedure is used to search for
dynamics of the fluctuations. For each event the ordered pseudorapidities,
$\e=-\ln\tan\frac{1}{2}\vt$  
($\vt$ is the polar angle of the particle), are scanned with a fixed
pseudorapidity window (bin) across the full $\e$-range of the event, and
the spike with $\dn$ number of tracks, hit in the window $\de$,
is determined.  Then the centers of spikes,
$\ec=(1/\dn)\sum_{j=1}^{\dn}\e_j$,
are calculated for all events and the distribution in $\ec$ is
investigated to reveal dynamical correlations. The distribution with
respect to the maximum density fluctuations, defined as
$\rmx=\dn_{\mx}/\de$, where $\dn_{\mx}$ is the maximum number of particles
per spike in each event for the chosen $\de$, is analyzed as well in order
to obtain dynamical character of the fluctuations observed.

The conclusions about dynamical content of the fluctuations could be
affected by the dependence of such an analysis on the form of pseudorapidity
distribution and the fact that the method of maximum fluctuations deals with the
mostly populated regions of the distribution. To avoid these and to compare
the results from different experiments, the ``cumulative'' variable,

\be
\ef(\e)\; =
\int_{\e _{\mn}}^{\e}
\r(\e ')\rd \e ' \,
/
\int_{\e _{\mn}}^{\e _{\mx}} \r(\e ')\rd \e '\; ,
\la{nv}
\ee
with  the uniform spectrum $\ddf$  within the interval [0,1] was proposed to be
used \ct{fl}. The transformed variable (\ref{nv}) are usually considered in
studying intermittency \ct{rev1}, also utilized in our recent reports
\ct{mn,mg,pl4}. Note, pseudorapidity is argued to be the most suitable
variable to analyze correlations providing intermittent structure of high-energy
events \ct{dijmp}.

\section{Results and discussion}

\subsection{Spike-center distributions}

The pseudorapidity spike-center distributions for the different size
$\dep$-bins and for spikes of the different density are presented in Fig.
1.  The widths of the bins are chosen to be compatible with those used
earlier \ct{y1,pl1}.  Multi-peak structure of the distributions can be
seen for $\dep=0.04$ ($\de\approx0.1$) (Fig. 1a) and 0.08 ($\approx0.2$)
(Fig. 1b).  However, one can observe two peaks placed in the region about
the same $\ecf$-positions with a tendency of the distributions to have a
double-peak shape as the size of the bin increases. The two peaks become
much more pronounced when the events are scanned by the large $\dep$, e.g.
of the width of 0.2 as shown in Fig. 1d.  Fitting these two bumps with
Gaussians, the peaks averaged over the different spikes, are found to be
placed at
$0.17$
and
$0.57$.
Recounted to the $\e$-variable, the positions of the peaks are of the values of
$0.60\pm0.05\st\pm0.12\sy$
and
$1.30\pm0.03\st\pm0.10\sy$
with the distance,
\be
d_0=0.68\pm0.06\st\pm0.16\sy
\la{d}
\ee
between them.

Recently, the study of the spike-centers has been carried out for hadronic
interactions at 205-306 GeV/$c$ \ct{na2}. The double-peak shape of the
$\ec$-distribution for pp-collisions vs. a single-peak structure in
$\pi$/Kp-interactions have been observed  in agreement with the coherent
gluon-jet emission model \ct{revd}.  The peaks has been found to be separated
by the distance of
$0.57\pm0.03\st\pm0.12\sy$,
also consistent with the model prediction. The double-peak form obtained here
for the central nuclear interactions are similar to that for pp-type reaction,
indicating superposition of nucleon-nucleon interactions in nucleus-nucleus one.
Moreover, the value of the distance found exceeds that in hadronic collisions
being in agreement with theoretical expectations \cite{dp}.

To observe dynamical correlation effect in these distributions, the analogous
distributions have been obtained from the simulated pseudorapidity one-particle
spectrum $\ddf$, in which, evidently, any information of two or more
multiparticle correlations is lost. The simulation procedure was as follows.  In
accordance with  the multiplicity  distribution of the experimental sample we
have randomly generated corresponding number of tracks.  Then  the
pseudorapidities  has been distributed according to the real $\ddf$-spectrum in
a quantity of the generated multiplicity. The total number of the events
simulated was 66300, so that it exceeded the experimental statistics by a factor
of 100. It is clear that the statistical properties of this set are completely
analogous to those of the ensemble resulted from arbitrary mixing of tracks from
different events, subject to the condition of retention of the
$\ddf$-distribution, and the obtained sample represents the result of
independent particle emission hypothesis.

The $\ecf$-distributions of the simulated events are shown in Fig. 1 by
open circles. Remarkable difference one can observe between the
experimental distributions (solid circles) and those obtained in
assumption of completely uncorrelated particle production. No whatsoever
peaks are seen in the latter case, following the background level and
manifesting the double-peak structure to be the dominant one. 

From the comparison of the experimental and simulated distributions in the
spike-centers proceeded for various $\dep$-windows and different many-particle
spikes, one can apparently conclude about dynamical character of the production
of high-density fluctuations. The dynamics of the intermittent structure of the
data studied could be assigned to the model of coherent gluon radiation
\ct{revd} as to one of a real candidate of the mechanism of formation of the
fluctuations.

To assess the reliability of the above conclusions and the stability of the
results we vary the investigation with changing the $\De$-range under
consideration as well as the polar angle, $\vt$, within the experimental error
$\al\ve_{\vt}\ar$. The observed character of the distributions remained
unchanged.

\subsection{Maximum density fluctuations}

The further analysis deals with the normalized distributions,
$(1/N){\rd}N/\rd\rmx$, of $N$ events as shown in Fig. 2 for four different
scanning windows.

The exponential decrease at $\rmx > \rmxa$ of these distributions, averaged over
all multiplicities $n$, is analogous to that observed in our previous studies
\ct{y1,pl1} and in other investigations performed for different  reactions
\ct{mx,mx1}.  Such a behavior differs from the poissonian one, expected for
processes with weak correlations of produced hadrons or for models of
multiperipheral or Regge types taking into account a limited number of reggeons.
The exponential behavior of the $\rmx$-spectra is argued to be a consequence of
primordial multiparticle correlations which are irreducible to two-particle
correlations \ct{rev4}.  The given inequality between the dispersion and the
mean values $\rmxa$ confirms the non-poissonian character of the distributions,
pointing out significant contribution of multi-particle correlations to the
local fluctuations observed.

Although the transition to the ``cumulative'' variable did not influence
much the above conclusion compared to the earlier results \ct{y1,pl1}, it
seems to be essential in studying the distributions at high $\rmx$-values.
Indeed, if the bell-like shape for small bins (Fig. 2a) is similar to that
observed in the previous reports, an increase of the $\dep$-width (Fig. 
2b, c) makes large $\rmx$-tails to appear in the $\rmx$-spectra. 

A change of the shape of the maximum density distribution from the exponential
to  more flat one with increase of the $\rmx$ seems to be in agreement with
that expected from the one-dimensional intermittency model \ct{mxt}.  The key
feature of the model is an existence of two regimes in particle production
process, described as turbulent and laminar components, leading  to two maxima
in the $\rmx$-distributions.

It is worthwhile that the model makes predictions for  the $\rmx$-distributions
considered at given multiplicity $n$. Note, the fixed-$n$ distributions are
energy and reaction-type independent that allows increasing statistics by
compiling results from different experiments. A study of the fixed-$n$
distributions has been carried out for hadronic interactions and the
large-$\rmx$ tails have been observed for  $\de = 0.1$ \ct{na1}.

In Fig. 2 we present the $\rmx$-distributions for two different narrow
$n$-intervals (solid squares and triangles). An effect of deviation of the
$\rmx$-distribution from the exponential behavior at large maximal densities are
well seen and became more prominent compared to the case of mixed-$n$
distributions. Already at $\de\approx 0.1$ (Fig. 2a) the shape of the
distributions develops tails at $\rmx > \rmxa$, apparent also for larger
$\dep$-bins, $\dep= 0.12$ and 0.2, but only for high-multiplicity events, $\mb$
(triangles in Fig. 2b, c). 

For the widest $\dep$ shown, $\dep=0.4$ (Fig.
2d), one can  not see the difference in the behavior of the distributions for
fixed $n$-regions.  However, the distribution for all multiplicities shows
slight flattening at large $\rmx$, but, in our opinion,  this is rather due to
the averaging over all $n$'s available.

Similar to the mixed-$n$ spectra the fixed-$n$ interval distributions show
their non-poissonian character resulting to an inequality between the
dispersion and the $\rmxa$ and indicating contribution of multiparticle
correlations. 

To reveal dynamical correlation effect, the obtained distributions are
compared to those based on the above described sample of the simulated
events, in which no whatsoever dynamical correlations exist.  The resulted
distributions for the four bins studied are shown in Fig. 2 for total
multiplicity as well as for the two fixed-$n$ intervals. The values of
$\c2$ give a quite good fit between the data and the generated spectra
independent of the width of the bins or the multiplicity. From this
comparison one can not conclude about (intermittency)  dynamics behind
observed multiparticle fluctuations. It seems that the dynamical
correlations in the distributions of maximum densities are too suppressed
by statistical ``noise'' to be detected.

However, it is worth to notice that the $\rmx$-distributions at $\dep=
0.12$ and 0.2 and $\mb$ (fig. 2b, c), obtained from the uncorrelated
emission hypothesis deviate from the data-based spectra at high-$\rmx$
values: if the flattening is well-seen for the measured distributions,
there is no change in the behavior of these distributions from the
simulated sample. 

The non-poissonian behavior found to be stronger for smaller bin-sizes and
the deviations between the data and the simulated distributions observed
for higher maximum densities, both seem to be a reason of non-statistical
fluctuations found by us via the method of normalized scaled factorial
moments \ct{mn,mg,pl3,pl4} and leading to extreme fluctuations recently
proposed to be searched for \ct{ext}. 

All these indicates a need of more detailed analysis to be done with
higher multiplicity reactions.  Note, that such an analysis could be
compared with that performed here due to the ``cumulative'' variable used
and fixed-$n$ intervals considered. 

A study of    influence of the error $\al\ve_{\vt}\ar$ in the measurement of
the polar angle $\vt$ of the produced charge particles demonstrated stability
of the obtained distributions and, therefore, the reliability of the
conclusions
above.

\section{Conclusions}

In summary, a study of spike production in central C-Cu collisions at 4.5
GeV/$c$ per nucleon is presented. To avoid the dependence of the results
on the form of the pseudorapidity distribution the transformation to the
uniform spectrum variable is utilized. The spike-center distributions and
those in maximum density fluctuations are investigated for various narrow
pseudorapidity windows to obtain dynamical collective effects. 

The double-peak shape of the spike-center distributions is found, similar to
that recently obtained in high-energy pp-interactions. The distance between the
positions of the peaks seems to be in agreement with the expectation of the
coherent gluon jet-emission model. Comparison with the results of completely
uncorrelated particle-production model  confirms dynamical origin of the effects
observed.

The maximum density distributions are investigated for mixed-$n$ values as
well as for fixed-$n$ intervals.  Non-poissonian character of the
distributions is found indicating contribution of multiparticle
correlations to the spikes.  A flattening of the shape of the
distributions at large maximum densities is obtained for high
multiplicities in qualitative agreement with one-dimensional intermittency
model.  The dynamics are found to be hidden by strong statistical
correlations, though visible deviations in the behavior of the simulated
distributions compared to the measured ones are indicated. 

To conclude, a direct study of two different approaches -- coherent vs. 
stochastic -- to the formation of the local dynamical fluctuations in
multiparticle production in central nuclear collisions at intermediate
energy is performed.  The coherent nature of dense group production is
found to be clearly manifested with a structure similar to that recently
found in hadronic interactions. Though no dynamics is obtained by means of
the stochastic (intermittent)  approach, the features of the maximum
density distributions indicate a possible origin of self-similarity of the
local fluctuations.

\vs*{1.2cm}

\noindent{\Large \bf Acknowledgements}\\

\noindent
We are grateful to the members of the GIBS (SKM-200) Collaboration for providing
us with the film data. Special thanks go to G.G. Taran for his kind assistance
in the data  handling and to B.A. Kulakov and V. Bradnova for their warm
hospitality.
We are  thankful to I.M. Dremin for his illuminating discussions. The useful
comments from all our colleagues have helped during the study and are
acknowledged. L.G. is thankful to the Russian Fund for Fundamental
Research for supporting her work under grant 96-02-19359a.

\np

\np
%%%%%%%%%%%%%%%%%%%%%%
% Figure captions    %
%%%%%%%%%%%%%%%%%%%%%%
\vs*{2.cm}

{\bf \large Figure captions}
\vs*{0.7cm}

\noindent
Fig.~1. Experimental ($\bullet$) and simulated ($\circ$)
spike-center distributions for  different $\dep$-bins and 
various $\dn$-number of particles per bin:

(a) $\dep\, = 0.04,\; \dn=4$,
(b) $\dep\, = 0.08,\; \dn=5$,

(c) $\dep\, = 0.12,\; \dn=7$,
(d) $\dep\, =  0.2,\; \dn=9$.

\vs*{0.7cm}

Fig.~2. Normalized experimental (solid symbols) and simulated (open symbols)
distributions in maximum density $\rmx$ for  different $\dep$-windows and 
three multiplicity patterns:

\noindent
(a) $\dep\, =0.04$, $\c2\se 1.3$ (\aln, $\rmxa\se 39.4$, $\rms \se 11.2$),

   $\c2\se 0.5$ ($\ma ,\; \rmxa\se 38.6$, $\rms \se 8.1$),

   $\c2\se 0.7$ ($\mb ,\; \rmxa\se 49.2$, $\rms \se 9.0$),

\noindent
(b) $\dep\, =0.12$, $\c2\se 1.2$ (\aln, $\rmxa\se 22.0$, $\rms \se 6.6$),

   $\c2\se 1.3$ ($\ma ,\; \rmxa\se 12.8$, $\rms \se 3.6$),

   $\c2\se 2.0$ ($\mb ,\; \rmxa\se 25.2$, $\rms \se 4.4$),

\noindent
(c) $\dep\, =0.2$, $\c2\se 1.1$ (\aln, $\rmxa\se 17.4$, $\rms \se 5.2$),

   $\c2\se 1.0$ ($\ma ,\; \rmxa\se 14.0$, $\rms \se 2.6$),

   $\c2\se 1.7$ ($\mb ,\; \rmxa\se 20.2$, $\rms \se 2.8$),

\noindent
(d) $\dep\, =0.4$, $\c2\se 0.9$ (\aln, $\rmxa\se 13.2$, $\rms \se 4.2$).

   $\c2\se 0.7$ ($\ma ,\; \rmxa\se 10.3$, $\rms \se 1.6$),

   $\c2\se 0.9$ ($\mb ,\; \rmxa\se 18.2$, $\rms \se 1.8$),

%%%%%%%%%%%%%%%%%%%%%%
%     Figures        %
%%%%%%%%%%%%%%%%%%%%%%

%\pagestyle{empty}

\np
\vspace*{1.5cm}

\begin{tabular}{l}
\epsfig{file=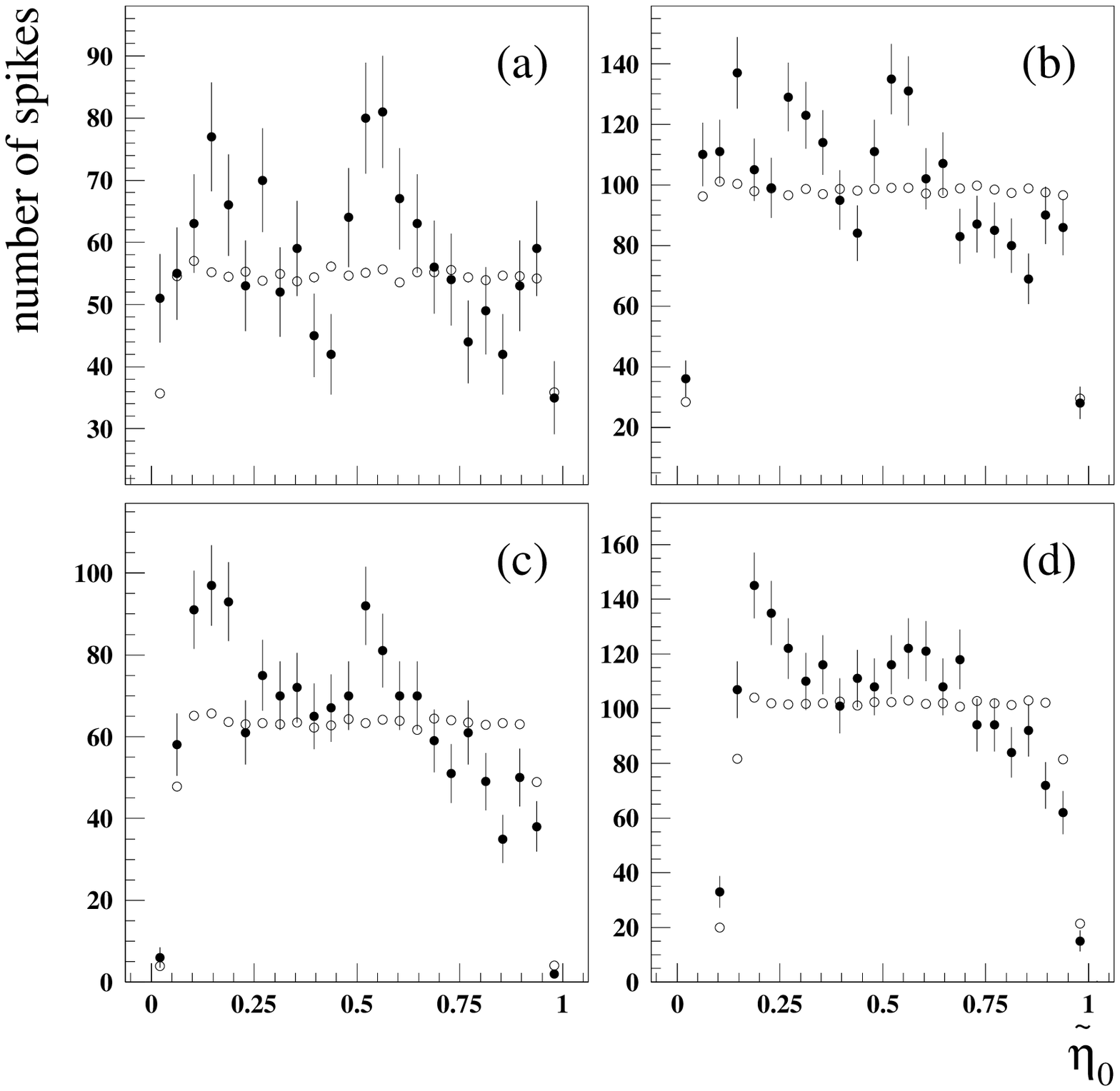,height=5.6in}
\end{tabular}

\vspace*{1.3cm}
{\centerline {\large Figure 1}}

\np
\vspace*{1.5cm}

\begin{tabular}{l}
\epsfig{file=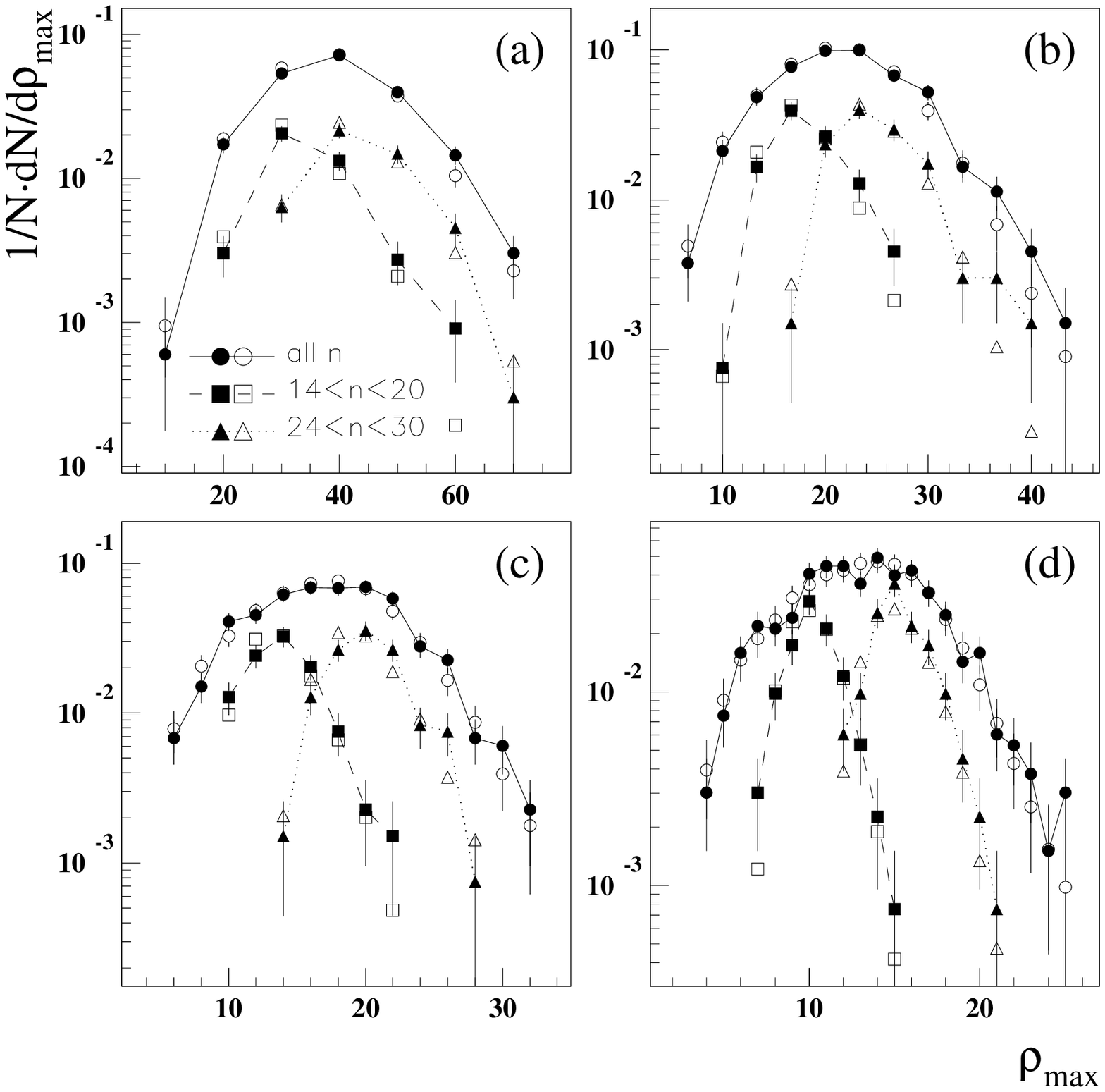,height=5.6in}
\end{tabular}

\vspace*{1.3cm}
{\centerline {\large Figure 2}}

\end{document}